\documentclass{aastex62}

\usepackage{lineno}
\usepackage{graphicx}
\usepackage{amsmath}
\usepackage{float}
\usepackage[CaptionAfterwards]{fltpage}
\usepackage{color}

\begin{document}

%\linenumbers

\title{Encounter of Parker Solar Probe and a Comet-like Object During Their Perihelia: Model Predictions and Measurements}

\correspondingauthor{Jiansen He}
\email{jshept@pku.edu.cn} 

\author[0000-0001-8179-417X]{Jiansen He}
\affiliation{School of Earth and Space Sciences, Peking University, \\
Beijing, 100871, P. R. China}

\author{Bo Cui}
\affiliation{School of Earth and Space Sciences, Peking University, \\
Beijing, 100871, P. R. China}

\author{Liping Yang}
\affiliation{State Key Laboratory of Space Weather, National Space Science Center, Chinese Academy of Sciences, \\
Beijing, 100190, P. R. China}

\author{Chuanpeng Hou}
\affiliation{School of Earth and Space Sciences, Peking University, \\
Beijing, 100871, P. R. China}

\author{Lei Zhang}
\affiliation{State Key Laboratory of Space Weather, National Space Science Center, Chinese Academy of Sciences, \\
Beijing, 100190, P. R. China}

\author{Wing-Huen Ip}
\affiliation{Institute of Astronomy, National Central University, Taoyuan, 32001, Taiwan, R.O. China}

\author{Yingdong Jia}
\affiliation{IGPP and EPSS, University of California, Los Angeles, CA 90095, USA}

\author[0000-0002-8990-094X]{Chuanfei Dong}
\affiliation{Department of Astrophysical Sciences, Princeton University, Princeton, NJ 08544, USA}

\author{Die Duan}
\affiliation{School of Earth and Space Sciences, Peking University, \\
Beijing, 100871, P. R. China}
\affiliation{Space Sciences Laboratory, University of California, Berkeley, CA 94720-7450, USA}

\author{Qiugang Zong}
\affiliation{School of Earth and Space Sciences, Peking University, \\
Beijing, 100871, P. R. China}

\author[0000-0002-1989-3596]{Stuart D. Bale}
\affil{Physics Department, University of California, Berkeley, CA 94720-7300, USA}
\affil{Space Sciences Laboratory, University of California, Berkeley, CA 94720-7450, USA}
\affil{The Blackett Laboratory, Imperial College London, London, SW7 2AZ, UK}
\affil{School of Physics and Astronomy, Queen Mary University of London, London E1 4NS, UK}

\author[0000-0002-1573-7457]{Marc Pulupa}
\affiliation{Space Sciences Laboratory, University of California, Berkeley, CA 94720-7450, USA}

\author[0000-0002-0675-7907]{John W. Bonnell}
\affil{Space Sciences Laboratory, University of California, Berkeley, CA 94720-7450, USA}

\author[0000-0002-4401-0943]{Thierry {Dudok de Wit}}
\affiliation{LPC2E, CNRS and University of Orl\'eans, Orl\'eans, France}

\author[0000-0003-0420-3633]{Keith Goetz}
\affiliation{School of Physics and Astronomy, University of Minnesota, Minneapolis, MN 55455, USA}

\author[0000-0002-6938-0166]{Peter R. Harvey}
\affiliation{Space Sciences Laboratory, University of California, Berkeley, CA 94720-7450, USA}

\author[0000-0003-3112-4201]{Robert J. MacDowall}
\affiliation{Solar System Exploration Division, NASA/Goddard Space Flight Center, Greenbelt, MD, 20771}

\author[0000-0003-1191-1558]{David M. Malaspina}
\affiliation{Laboratory for Atmospheric and Space Physics, University of Colorado, Boulder, CO 80303, USA}

\begin{abstract}

Parker Solar Probe (PSP) aims at exploring the nascent solar wind close to the Sun. Meanwhile, PSP is also expected to encounter small objects like comets and asteroids. In this work, we survey the ephemerides to find a chance of recent encounter, and then model the interaction between released dusty plasmas and solar wind plasmas. On 2019 September 2, a comet-like object 322P/SOHO just passed its perihelion flying to a heliocentric distance of 0.12 au, and swept by PSP at a relative distance as close as 0.025 au. We present the dynamics of dust particles released from 322P, forming a curved dust tail. Along the PSP path in the simulated inner heliosphere, the states of plasma and magnetic field are sampled and illustrated, with the magnetic field sequences from simulation results being compared directly with the in-situ measurements from PSP. Through comparison, we suggest that 322P might be at a deficient activity level releasing limited dusty plasmas during its way to becoming a ``rock comet''. We also present images of solar wind streamers as recorded by WISPR, showing an indication of dust bombardment for the images superposed with messy trails. We observe from LASCO coronagraph that 322P was transiting from a dimming region to a relatively bright streamer during its perihelion passage, and simulate to confirm that 322P was flying from relatively faster to slower solar wind streams, modifying local plasma states of the streams.

\end{abstract}

\keywords{minor planets, asteroids: general; comets: general; (Sun:) solar wind}

\section{Introduction}

Comets and comet-like objects as a kind of messenger and fossil of solar system formation are distributed widely in the vast space of solar system, with some of their perihelia and aphelia extending close to the Sun and the solar system outer boundary, respectively. Previously, several space missions (e.g., ICE, Giotto, and Rosetta) had been implemented in the form of flyby, orbiting, landing, or impact. The International Cometary Explorer (ICE) was the first-ever comet encounter, passing through the tail of Comet Giacobini-Zinner in 1985 within a distance of 7860 km from the nucleus \citep{Smith1986}, and then flew through the tail of Comet Halley in 1986 at a larger distance (31 million km) during its closest approach \citep{Stelzried1986}. Giotto took the first ever closest image of a comet nucleus at an approach distance of less than 600 km from Comet Halley’s nucleus, showing ejection of gas and dust into space from some active regions \citep{Keller1986}. Upstream wave activities of comet origin was discovered by Giotto at about 2.8 solar radii (Rs) (0.013~au) away from Comet Halley’s nucleus \citep{Neubauer1986}. The surface, subsurface, structure, and ejection activity of 67P have been studied in detail \citep{Sierks2015, Rotundi2015, Gulkis2015, Auster2015}. 

During the past thirty years, various plasma interaction models (using magnetohydrodynamics(MHD), multi-fluids, hybrid, full-particles) have been conducted to study cometary interactions with the background solar wind \citep{Jia2008, Gortsas2009, Rubin2014, Koenders2016, Deca2017}. Using MHD simulations, plasma structures (e.g., bow shock and plasma tail) and their dynamics have been investigated under different solar wind conditions at various heliocentric distances \citep[e.g.,][]{Gombosi1996, Hansen2007, Jia2009}. Multi-fluid model is applied to study the heavy ion dynamics in the interaction between sun-grazing comet and solar corona, reproducing the tail of $O^{\rm{6+}}$ consistent with the observations in extreme ultraviolet channel \citep{Jia2014}. Numerous mass loading processes (e.g., photo-ionization, electron impact ionization, charge exchange) are also incorporated in multi-fluid model, displaying that the illumination-driven neutral gas outflow can lead to the formation of magnetic reconnection and nucleus-directed plasma flow inside the night-side reconnection region \citep{Huang2016}. The multi-fluid models are usually limited in the local 3D simulations, which set the fixed upstream solar wind conditions as the boundary conditions of one side. Using global 3D simulations, the MHD model with two species flowing in one fluid is employed to study the evolving cometary structures and dynamic solar wind disturbances as a consequence of the comet’s orbital passage through the background solar wind \citep{Rasca2014}. The inner spherical boundary condition of magnetic field was simply assumed to be a dipole in \citet{Rasca2014}. Our work in this paper advances to incorporate the observed global magnetogram into the inner spherical boundary conditions.

The comets encountered/orbited by spacecraft were beyond the Venus’s orbit. Since the era of Solar and Heliosphere Observatory (SOHO), more sun-grazing comets, sunskirter and near-Sun comets have been discovered by using spaceborne coronagraphs like Large Angle and Spectrometric Coronagraph (LASCO) onboard SOHO \citep{Biesecker2002, Knight2010, Lamy2013, Battams2017}. Please refer to \citet{Jones2018} for a comprehensive review on this subject. The remote observation of near-Sun comets was usually taken at a distance of 1 au, and there is no dedicated plan to conduct a close observation of near-Sun comets in the near future. Chances of occasional encounters between distant cometary tails and PSP as well as Solar Orbiter in the inner heliosphere are anticipated, e.g., the possible ever-cross of Atlas's ion and dust tails by Solar Orbiter in early June of 2020 \citep{Jones2020}. The DESTINY+ mission to be launched in the next few years is going to target and flyby the active asteroid 3200 Phaethon and study its surface as well as dust release \citep{Arai2018}. The activity of a comet-like object near the Sun depends on both the external environment and its own property. The external environment changes a lot with distance decreasing from 1 au to 0.2 au: leading to the exposure to both more intense solar irradiation and stronger solar wind impact by a factor of 25 as a first-order approximation, under the reasonable assumption of constant solar wind speed and $r^{-2}$ dependence of number density between 0.2 and 1 au. The volatile icy and refractory rocky materials on the comet-like body are likely to be sublimated, melted, and evaporated as they become hotter when getting close to the perihelion within 0.2 au \citep{Mann2004}. The released materials are believed to play a crucial role of replenishing dust particles and pick-up ions into the inner heliosphere \citep{Geiss1996}. In order to closely observe and measure a comet-like object at its most active time near perihelion, the observer has to travel close to both the Sun and the small object. Given PSP’s proximity to the Sun, it offers a unique opportunity to observe near-Sun small object's activity closely. As the mission goes, we expect more encounter cases for such investigations.

As one of the first short-period (3.99 year period) “comets” with near-Sun orbit discovered via SOHO, which has recorded more than 3000 comets, the trajectory of 322P/SOHO (322P for short) indicates that it does not belong to the traditional near-Sun cometary groups (e.g., Kreutz group) but regarded as a sporadic near-Sun object \citep{Honig2006}. It is still a hot debate on the origin of 322P, e.g., cometary origin or asteroid origin. The Tisserand parameter ($T_{\rm{J}}$) of 322P is calculated to be 2.3, falling in the range of [2, 3] for Jupiter-family comets, while being less than the lower threshold of asteroid ($T_{\rm{J}}>3$) \citep{Levison1996}. 322P exhibits asymmetric light-curves at each apparition, which is common on comets, but 322P does not show obvious tail or coma in the field of view of LASCO-C2 onboard SOHO \citep{Lamy2013}.  The low activity of 322P during its perihelion implies that 322P, if of cometary origin, may be a nearly bared comet with few icy materials for sublimation. The asymmetric light curve implies 322P might still have a large but unresolved cross-section of dust. Please note that, the resolution of LASCO is so low that any cometary feature if existent may as well just go unresolved. The small body ejecting dust (“dust” refer to “fine particles of matter”), e.g., (3200) Phaethon, was recently named “rock comet” to be distinguished from the conventional comets and asteroids. Phaethon is the body responsible for the Geminid shower in December, and is revealed to extend a comet-like tail when flying around the perihelion (~0.14 au) where the temperature can be as high as 1000 K \citep{Jewitt2013}. The rock of Phaethon is speculated to be heated, thermal disintegrated, fractured, and finally turned into dust particles for ejection. Those hypothetical activities of the so-called “rock comet” have never been observed and confirmed at a close distance.

\section{Possibility and Example of Encounter between PSP and Near-Sun Small Object}

With the orbital elements of comets from “cometels.json” provided by Minor Planet Center (MPC) of International Astronomical Union (IAU), we calculate the preliminary ephemeris of every comet in catalog, and compare them with the PSP’s ephemeris during the 7 years of nominal mission phase of PSP. The ephemerides of possible encounter candidate comet are double-checked with the JPL/Horizons database. We find that, PSP would fly by the comet-like object named "322P" at the nearest distance of about 0.025 au (5.4 Rs) around 21:30 UT on 2019 September 2 (2019-09-02 as short format “yyyy-mm-dd” and used hereafter). The ephemerides calculated from MPC/IAU and downloaded from JPL/Horizons for 322P are basically consistent with each other. The 322P perihelion happened around 09:00 UT on August 31 at the heliocentric distance of 0.054 au. Two days later, 322P was encountered by PSP at their closest distance. So this encounter took place during the outbound of 322P, and we could expect more dust release as compared with its inbound.

The diameter of 322P is inferred to be 150-320 m based on its high albedo value (0.09-0.42) \citep{Knight2016}. Moreover, the color and density of its nucleus make it distinguished from usual comet nucleus, suggesting the possibility of asteroid origin \citep{Knight2016}. If 322P is an asteroid, then its activity may be due to sublimation of some refractory materials like silicates \citep{Kimura2002} and dust release via non-traditional means like thermal fracture rather than outgassing drag. The near Sun orbits including sungrazing orbits have been considered to be the ultimate orbital state of many main-belt and near-Earth asteroids \citep{Gladman1997}. Therefore, 322P is clearly not a typical comet, but in this paper we are going to call it a comet-like object for simplicity. The encounter of 322P by PSP may provide a timely unique opportunity to decipher the eruption mechanism and origin of this object.

\section{Dynamics of Dust Released from 322P}

The ephemerides of PSP and 322P during their encounter are plotted in the Heliocentric Ecliptic Coordinate system in Figure 1a. The time interval of display is from 2019-08-03 to 2019-10-03. It can be seen that, on 2019-09-02,  about 3 days after the perihelion, 322P gets closest to PSP, which is also right after the perihelion. 

The dynamics of dust is important to the environment around 322P. Different forces may govern on different sizes of particles \citep{Ragot2003}. According to \citet{Morfill1986}, the relative importance of radiation pressure depends on two factors: one is its light absorption, and the other is its size. The absorbing particles (e.g., carbon and magnetite) usually sense higher radiation pressure than the dielectric particles (e.g., silicate). For dust particles larger than a sub-micrometer (r$>$0.1 $\rm{\mu m}$), it is the radiation pressure ($\mathbf{F}_{\rm{rad}}$) and the solar gravity ($\mathbf{F}_{\rm{grav}}$) that are controlling their movements. The ratio between radiation pressure and solar gravity ($\beta=F_{\mathrm{rad}}/F_{\mathrm{grav}}$) is inversely proportional to the particle size, and reaches to a maximum level for the size between 0.1~$\rm{\mu m}$ and 1~$\rm{\mu m}$. For small dust particles with sizes less than the sub-micrometer scale (r$<$0.1 $\rm{\mu m}$), radiation pressure becomes less important again as the particle size reduces to the sub-micrometer level according to \citet{Mann2010}. The force ratio $\beta$ can be as small as 0.1 when particle size is at the nano-meter level \citep{Ip2012}. The Lorentz force ($\mathbf{F}_{\rm{EM}}$) and solar gravity force ($\mathbf{F}_{\rm{grav}}$) are responsible for the dynamics of such small dust particles. Large dust particles form a tail carrying charges without being significantly affected by the solar wind electromagnetic fields, so they can be studied as test particles released along the comet orbit and subjected to the sum of two forces ($ m\frac{d\mathbf{V}}{dt}=\mathbf{F}_{\mathrm{rad}}+\mathbf{F}_{\mathrm{grav}}$). Dust particles exposed to different force ratios ($\beta$=0.003, 0.03, 0.1, 0.2, 0.4, 0.6, and 0.8) are shown with different colors in Figure 1b-1e, which exhibit the snapshots at four release times. The syndynes after the perihelion curve clockwise, quite different from that before the perihelion, that look similar to the trajectory of 322P. The clockwise curved shape of the syndynes is due to the blowing away of dust particles by solar radiation during the perihelion. For larger grains (e.g., $\beta$=0.01 or smaller), their dynamics is mainly controlled by the solar gravity and therefore travelling in orbits similar to that of 322P. The twist segment on the syndyne curves of dust particles (see the brown and yellow lines in Figure 1c for example) is caused by two factors: (1) abrupt change of force on dust particles before and after their release from 322P; (2) force change near the perihelion phase. The dust particles with higher $\beta$ values have their twist segment taking place earlier than their counterparts with lower $\beta$ values. As time goes on, e.g., from September 02 (Figure 1c) to September 13 (Figure 1e), the small twist segments develop into large curved segments. We can also see from Figure 1c-1e that, the dust particles with extreme low $\beta$ values ($\beta < 0.1$) move around their mother body 322P instead of being thrown away from 322P like the other dust particles, on which radiation pressure dominates. We note that, although the particles up to mm-sized with extreme low $\beta$ ($\beta<0.1$) will not decouple significantly from their parent 322P at the beginning of their release, but their deviation from the parent’s orbit will become evident after a certain period of flight especially during the perihelion phase.

The closest distance between PSP and 322P is found to be less than 6 Rs. The spatial resolution of images would therefore be enhanced about 35 times when observed from PSP as compared to the view point from the Earth. This would be helpful if 322P and its tail fall within the field-of-view (FOV) of WISPR (Wide-field Imager for Solar Probe Plus) onboard PSP \citep{Vourlidas2016}. Unfortunately, the 322P nucleus is outside the FOV of WISPR during the encounter (see Figure 1f-1i). WISPR is pointing to the right of the Sun, while 322P is located on the left side of the Sun, which could be captured by a hypothetical camera only if it is pointing in the direction that mirrors that of WISPR. The trails of the dust particles with different $\beta$ values from 0.1 to 0.8 are also illustrated in Figure 1f-1i, which show highly curved profiles of the dusts’ footprints in the projected plane of sky.  The time varying distances between dust particles and PSP can be less than 1 Rs at its shortest distance during [12:00, 16:00] UT on 2019-09-02. Different $\beta$ values means different ratios between $F_{\rm{rad}}$ and $F_{\rm{grav}}$. For particles of larger $\beta$ value, they would sense relatively more anti-sunward radiation pressure force and hence deviate more away from the orbit of nucleus. The optical visibility of dust particles is determined by the following factors: (1) the distance between the Sun and the dust particle, (2) the distance between the dust particle and the observer, (3) the angle between the vector from the Sun to the particle and the vector from the particle to the observer, (4) the characteristics of the dust particle (e.g., size and scattering cross-section) (see Chapter 5 in \citet{Schwenn2012} for details). Different $\beta$ values might affect the visibility of different particles through the above four factors.

The dust particles would be charged and reduced in size by ablation down to nanograins, and then be accelerated by the electromagnetic field of the solar wind. The dust particles, if impact into the spacecraft at high speed, can generate plasma cloud causing electric field spikes detectable by onboard electric antennas \citep{Meyer2009, Ip2012, Juhasz2013}. It is possible that PSP would be impacted more by the dusts released from 322P during their encounter than other times. The quantitative estimate of how many dust impacts PSP could sense is another topic beyond the scope of this work, which requires knowledge of the electric antennas’ performance when measuring the signal generated by dust impact, as well as the cross-section of the spacecraft along the direction of dust impact \citep{Mozer2019, Szalay2019}. A constraint on the dust production of 322P could be placed based on the negative or positive detections of dust impacts. From the simple dynamic modeling result of released dust particles considering only solar gravity and radiation pressure, we know that the dust particles may approach the PSP spacecraft in bulk around September 02 (see Figure 2). The closest encounter between dust particles and PSP with the shortest distance of less than 0.2 Rs is calculated to take place at around 08:00~UT on September 2. The dust particles responsible for the closest encounter are found to be released from 322P about two days before their encounter with PSP. We survey the time domain samplers at a cadence of 7s, and find about 1353 samples with electric potential difference between antenna tips being larger than 50 mV on September 02. Dust strikes may be playing an important role in these events. In order to distinguish the three types of candidates for short duration spikes in electric field (dust impact, time domain structures, and turbulence), more investigation is needed, which is beyond the scope of the present work.

It is interesting to learn that WISPR onboard PSP observed the dust trail following the orbit of (3200) Phaethon during its first solar encounter in November 2018 \citep{Battams2020}. The analysis result of \citet{Battams2020} implies that the observed dust trail may contribute partially to the Geminid meteor stream through its perihelion and may be more than recent release due to perihelion activity of (3200) Phaethon. FIELDS onboard PSP can play as the synergy with WISPR. The magnetic and electric field antennas of the FIELDS instrument suite could be used to investigate possible changes in the wave dispersion relation and wave propagation in case that the spacecraft is flying through a dusty plasma. The level-3 data during September 02 and 08 due to a data corruption are not available on the WISPR website. Only the images taken before the encounter between PSP and 322P are displayed here in Figure 3. From Figure 3a and 3b, which were taken by WISPR on August 29 and at an earlier time on August 31, we can clearly see the coronal streamers extending into the inner heliosphere. While the messy trails in Figure 3c and 3d, which were photographed by WISPR at 07:00 and 23:00 UT on August 31, are likely spallation products from dust hits onto the PSP spacecraft structure.

During its perihelion passage, 322P was observed by LASCO C2 and C3 in the time intervals of [2019-08-30 23:12, 2019-08-31 03:36] UT and [2019-08-31 03:42, 2019-09-01 05:54] UT, respectively. The time sequences of 322P’s traces in the field-of-views (FOVs) of C2 and C3 are illustrated in Figure 4a and 4b. Similar to most previous reported near-Sun comets/asteroids, there seems to be no significant signal of dust tail or gas tail extending from the 322P itself. The gradual brightening of 322P as it approached to the Sun is clearly identified in Figure 4a. 322P was transiting from left to right in the FOV of C2 during its inbound towards its perihelion. In the larger FOV of C3, 322P’s trace shows a switch-back curve (see Figure 4b). After the switch-back, 322P moved further away from the Sun as well as LASCO/SOHO, meanwhile, it flew into a bright solar wind streamer belt at higher latitude as shown in Figure 4b. As a consequence, 322P looks fading away and is gradually obscured by the streamer belt in the images recorded by C3. The traveling towards the streamer belt (usually characterized with higher density and lower speed) during its post-perihelion phase is also successfully reproduced in our numerical modeling results, which will be presented hereafter. In the projected plane from the perspective of LASCO, 322P is closest to PSP around 01:00 UT on 2019-09-01 (see the center of green circle and gray diamond in Figure 4b for their projected positions). Their real closest encounter in the 3D space is calculated to be around 21:30 UT on 2019-09-02.

\section{Mass-loading of Released Mass onto Near-Sun Solar Wind}

If the dust particles released from 322P near the Sun have nanograins with their sizes being smaller than 10 nm, then similar to the force comparison result done by \citet{Ragot2003}, the dominant force term would be electromagnetic force, which is larger than the second dominant force term (gravitational force) by two orders of magnitude and far more larger than the other force terms (e.g., radiation pressure force). Such magnitude rank of force terms indicates that the nanograins and even smaller particles can be regarded as pickup charged particles in the solar wind, causing mass loading effect to the solar wind. The smaller particles (e.g., nanograins, charged molecular and atomic ions), which are susceptible to the solar wind electromagnetic field, if at sufficient density, can disturb the solar wind and interplanetary magnetic field. We simulate this process on a two-species plasma description, which introduces the released plasma (may be composed of heavy ions of refractory elements) and picked up instantly with ambient solar wind for simplicity, while ignoring the detailed finite Larmor radius (FRL) effect during the pickup process for a qualitative estimation. In order to set the inner boundary more conveniently, we simulate this interaction in the solar co-rotating reference system (i.e. Carrington coordinate system). Correspondingly, the ephemeris of 322P is also changed to Carrington coordinate system, and serves as the moving point source of mass release. 

In our model, a solar wind ion and a charged ejecta species of mass are simulated, while electrons keep the electric neutrality and contribute to the thermal pressure. Our two-species plasma model conserves mass, momentum, and energy using the set of equations adopted by \citet{Rasca2014}, which are given as followings,

\begin{linenomath*}
%\begin{subequations}
  \begin{eqnarray}
  \frac{\partial\rho_{\mathrm{SolarWind}}}{\partial t}+\nabla\cdot(\rho_{\mathrm{SolarWind}}\mathbf{u})=0,\label{eqn:RhoSolarWind}\\
  \frac{\partial\rho_{\mathrm{MassRelease}}}{\partial t}+\nabla\cdot(\rho_{\mathrm{MassRelease}}\mathbf{u})= S_\rho, \label{eqn:RhoMassRelease}\\
  \rho=\rho_{\mathrm{SolarWind}}+\rho_{\mathrm{MassRelease}}, \label{eqn:RhoTotal}\\
  \frac{\partial\rho\mathbf{u}}{\partial t}=\nabla\cdot[\rho\mathbf{uu}+(p+\frac{B^2}{2\mu_0})\mathbf{I}-\frac{\mathbf{BB}}{\mu_0}]=\mathbf{S}_{\rho\mathbf{u}}, \label{eqn:RhoU}\\
  \frac{\partial\mathbf{B}}{\partial t}+\nabla\cdot(\mathbf{uB}-\mathbf{Bu})=0, \label{eqn:dBdt}\\
  \frac{\partial E}{\partial t}+\nabla\cdot[\mathbf{u}(E+p+\frac{B^2}{2\mu_0})-\frac{(\mathbf{u}\cdot\mathbf{B})\mathbf{B}}{\mu_0}]=S_E, \label{eqn:dEdt}
    % \omega_j&=&v_{\mathrm{A}}\hat{\mathbf{B}_0}\cdot\mathbf{k}_j+\mathbf{v}_{\textrm{SW}}\cdot\mathbf{k}_j,\label{eqn:dispersion}\\
    % {\mathbf{B}}_j&\parallel&(\mathbf{B}_0\times\mathbf{k}_j).
  \end{eqnarray}
%\end{subequations}
\end{linenomath*}
where $\rho_{\rm{SolarWind}}$, $\rho_{\rm{MassRelease}}$, and $\rho$ represent respectively the mass densities of background solar wind, ejecta from small object, and mixture, $\mathbf{u}$ is the averaged plasma velocity vector, $E$ is the total energy density (bulk kinetic + thermal kinetic + magnetic), $S_\rho$, $S_{\rho\mathbf{u}}$, and $S_E$  represent the source terms of mass density, momentum density, and energy density. In specific, the term $S_\rho=\frac{\dot{M}(\mathbf{x}-\mathbf{x}_{\rm{MassRelease}})}{\Delta \mathrm{Vol}}$ denotes the local mass release rate per unit volume. The released small particles range from ions to charged nanograins of different mass, but we simplify them into one mass release rate for qualitative estimation of the tail. We are only concerned about mass-loading effect to the solar wind after the coupling. The momentum source $S_{\rho \mathbf{u}}=S_{\rho} \mathbf{u}_{\text {MassRelease}}-\rho\left[\frac{G M}{r^{3}} \mathbf{r}+\mathbf{\Omega} \times(\mathbf{\Omega} \times \mathbf{r})+2 \mathbf{\Omega} \times \mathbf{u}\right]$ contains the momentum change rate per unit volume contributed by the released mass,  $S_{\rho} \mathbf{u}_{\rm{MassRelease}}$, the solar gravity, and the inertial forces (centrifugal force and Coriolis force). The velocity,  $\mathbf{u}_{\rm{MassRelease}}$, is the velocity of released mass injecting into the simulation domain in the corotating reference frame. The energy source term, $S_{E}=\frac{1}{2} S_{\rho} u_{\text {MassRelesse }}^{2}-\rho \mathbf{u} \cdot\left[\frac{G M}{r^{3}} \mathbf{r}+\mathbf{\Omega} \times(\mathbf{\Omega} \times \mathbf{r})\right]+Q_{\text {heat}}$, comprises the bulk kinetic energy density increase per unit volume contributed from the newly released mass, the work done by the solar gravity and the centrifugal force, and the heating process in the solar corona and solar wind. 

We employ the “Conservation Element and Solution Element” (CESE) scheme to the six patches of spherical grids as described by \citet{Feng2012}, which is dedicated to study the physics of solar-interplanetary (SIP) space in their SIP-AMR-CESE MHD model. The initial conditions for the flowing plasmas are the Parker solution to the steady solar wind. The initial state of magnetic field is the potential field extrapolation result from the global magnetogram of CR2216. The inner boundary conditions are fixed values equal their initial condition, and rotates with the Sun. The outer boundary conditions are implemented with zero-gradient extrapolation. Adaptive mesh refinement (AMR) is applied to specific regions of interest with two criteria: grid is refined when current density is greater than the threshold, to capture the heliospheric current sheet; grid is also refined at places where the released dusty plasma is dense enough to capture the dusty plasma tail. 

Three levels of ejection activities (low, intermediate, and high) are experimented, with the mass release rates  $\dot{M}=2\times10^3$~kg/s, $2\times10^4$~kg/s, and $2\times10^5$~kg/s, respectively. \citet{Knight2016} provided an upper limit of 2000 kg/s based on the derivation from optical data, which may be most sensitive to multi-micron-sized dust. The upper limit of mass release rate for the general near-Sun asteroid/comet population is derived to be $\sim10^5$ kg/s by assuming that multi-micron-sized dust dominates ejecta \citep{Ye2019}. The mass release rate of sub-micron-sized dust, however, is unavailable from observations at present. Here, we assume that the maximum mass release rate of dusty plasmas may be approximated as the similar level. It is an open question on what proportion of the mass loss from a near-Sun comet is allocated to nano-dust particles, which are further charged to become dusty plasma and coupled with background solar wind. Spacecraft measurements of Halley comet infer a small contribution of nano-dust ($<6\%$) to the mass loss \citep{Mann2017}. However, Halley is not a near-Sun object, which may have dust particles to be more likely fragmented into small pieces (e.g., nanometer-scale sizes) near the Sun. Although the signs of dust fragmentation have been observed on several comets using ground-based observations \citep[e.g.,]{Combi1994, Schleicher2002}, the definite proportion of nano-dust to the total dust mass loss for near-Sun objects requires dedicated measurements in the future. 

In the modeling domain, along with the passage of 322P, mass is continuously released and lost from 322P at the designated rate. 322P acts as a moving point source of mass release, which is mass-loaded into the solar wind creating dusty plasma structure in the inner heliosphere. The simulation result is shown in Figure 5 with velocity and released dusty plasma density contours sliced in a specific plane, which is determined by three points: the Sun, the object nucleus, and the end of the tail contour. There are some similarities between the three cases: (1) the ejecta slows down the solar wind; (2) the ejecta forms a tail extending anti-sunward and curving clockwise (westward). The tail is shaped by two processes: (1) orbital motion of 322P relative to the solar wind flow; (2) the velocity of released dusty plasma as picked up by the solar wind. There also exist some evident differences between the three cases: (1) higher level of activity results in stronger turbulence and wider tail; (2) lower-level activity loads less mass onto the solar wind and therefore causes less deceleration, leading to a faster development of the tail.

In Figure~6, we illustrate the distributions of $B_{\rm{r}}$ (magnetic radial component) and $P_{\rm{th}}$ (plasma thermal pressure) as disturbed by the loaded mass. There is an enhancement and a reduction of original $B_{\rm{r}}$ surrounding the comet and its tail, respectively. Furthermore, in Figure 6b \& 6c, $B_{\rm{r}}$ shows bipolar stripes roughly aligned along the mass-loaded structure, indicating field line draping. Please note that $B_{\rm{r}}$ has been multiplied by a factor of 10 to compensate the weak open magnetic field strength when imposing the interface radius of Potential Field Source Surface (PFSS) at 2.5 Rs from solar center. The weak open magnetic field obtained from PFSS-extrapolation may be related with the underestimate of magnetic field strength of solar magnetogram at higher latitudes. The interplanetary magnetic field lines as strongly disturbed by the simulated cometary plasmas display a geometry with additional double kinks (switchback) in Figure 6c-1 (an inset in Figure 6c). In Figure 6e \& 6f, thermal pressure $P_{\rm{th}}$ is evidently enhanced at the head of the structure due to the compression after the shock. $P_{\rm{th}}$ on the west side decreases and forms an elongated rarefaction region as the wake of the comet. The modeling distributions of $B_{\rm{r}}$ and $P_{\rm{th}}$ provide the context of environment, through which PSP would pass and make measurements. The details of simulated tail crossing along the same trajectory as PSP together with the in-situ measurements of interplanetary magnetic field from PSP will be given in the next section.

\section{Simulated Variables Along the Path of  PSP's Trajectory}

Since 322P is speculated to release a limited amount of dust and gas when approaching its perihelion, we focus on analyzing the case of low activity. As noted in previous section, for the interaction between 322P released materials and background solar wind, we have assumed three levels of activities, which have different mass release rates for the low, middle, and high levels of activities, respectively. The mass release rate of the low-level activity is consistent with the constraint implied in \citet{Knight2016}. To highlight the evolution of dusty plasma tail and disturbed solar wind plasmas, in Figure 7, we display six snapshots of velocity distribution and isosurface of dusty plasma density at a certain level (e.g., 1~$\rm{amu/cm^3}$). The isosurface of dusty plasma density is observed to bend eastward of the radial direction. The lengthening trend of the tail is a time effect that more dusty plasma is released and mass-loaded into the background solar wind. Note that the plane of slice displaying the velocity distribution is determined by the same set of three points as we used in Figure 5, and has different inclination angles between snapshots. On the other hand, Figure 7 also shows that the time-varying background solar wind around 322P has changed from relatively fast stream to slow stream between 2019-09-01 and 2019-09-03, which may affect the real interaction quantitatively. The solar wind velocity distribution is plotted as variable $\mathbf{u}$, which is explicitly governed by Eq~(4). The formation of fast and slow solar wind streams is determined by their respective flow tube geometries (e.g., expansion and curvature) as well as their energy deposits caused by heating process along with their outflow journeys. The streamer belt as identified in Figure 4 where PSP entered after its perihelion is well consistent with the slow stream displayed in Figure 7e and 7f, which is the background solar wind for 322P during its outbound journey right after the perihelion. 

In the Carrington coordinates, we find that 322P flies much faster than the solar rotation at its perihelion, while PSP roughly co-rotates with the Sun around its perihelion. Both 322P and PSP approximately co-rotate with the Sun during their encounter. The trajectories of 322P and PSP, which are color coded with the rainbow table to denote the elapsed days centered around the time of 00:00 UT on 02/09/2019, are also plotted in Figure 7. Figure 7 depicts that PSP flies successively inside (23:24 UT on 2019-09-01), west side (09:47 UT and 13:56 UT on 2019-09-02), sunward/ahead (06:32 UT and 16:54 UT on 2019-09-03) of the dusty plasma tail.

If PSP passes through the dusty plasma tail, perturbations may be measured by its instruments, e.g., FIELDS for electromagnetic fields \citep{Bale2016}, SWEAP for plasma moments \citep{Kasper2016}. Typical examples of such perturbations are compared in Figure 8 between the three activity levels as modeled. The number densities of released and loaded mass, if only taken the H element into account ($N_{\text {MassRelease, H }}=\rho_{\text {MassRelease }} / m_{\mathrm{H}}$), can be as high as 10~$\rm{amu/cm^3}$ for the low-level activity. The values approach to the level of background solar wind number density ($N_{\text {SolarWind, H }}=\rho_{\text {SolarWind }} / m_{\mathrm{H}} \sim 200 \mathrm{cm}^{-3}$) for the middle-level activity, and exceed $N_{\rm{SolarWind,H}}$  by an order of magnitude for the high-level activity. Reductions in the flow velocity ($\mathbf{u}$)  are displayed in Figure 8b: a reduction of less than 20~km/s, more than 50~km/s, and more than 100~km/s for the low-level, middle-level and high-level activities, respectively. 

The different morphological presence of 322P when near the Sun versus classical active comets indicates that the extra electron number density associated with the released and charged particles from 322P is significantly smaller than the electron number density of background solar wind. As we can see in Figure~8a, the released densities of the three levels of mass release rates are less than the solar wind density ($\rho_{\rm{SolarWind}}$), comparable to $\rho_{\rm{SolarWind}}$, and larger than $\rho_{\rm{SolarWind}}$ but no more than two orders of magnitude. When assuming the charge-to-mass ratio of nanograins to be $10^{-5}$~$\rm{e/m_p}$ to calculate the electron number density ($N_{\rm{e,MassRelease}}$) associated with the charged nanograins, we find that they are all smaller than the electron number density of local solar wind ($N_{\rm{e,SolarWind}}$) by at least three orders of magnitude. Therefore, the intensity scattered by the released electrons is weaker than its counterpart by the solar wind electrons.

As we have discussed for Figure 6, the simulated radial component of magnetic field vector ($B_{\rm{r}}$) as sampled along the path of PSP should also change during the encounter. The polarity of $B_{\rm{r}}$ can be reversed from the sunward direction to the anti-sunward direction when the activity level is high enough. This kind of $B_{\rm{r}}$-reversal may cause the complexity of magnetic field measurement by PSP and corresponding interpretation (switch back / kink of magnetic field line). The time curves of $B_{\rm{r}}$ component (see Figure 8c), which is negative at the two asymptotic sides, exhibit a “W”-shape for all the three simulation cases: a dip in the front, a bump at the center, and a dip at the back. The two dips of the negative $B_{\rm{r}}$ corresponding to the enhancement of $|\mathbf{B}|$ on both sides indicate the piling-up of magnetic field flux on the flank regions as they drape around the obstructing 322P. On the other hand, the bump is caused by the magnetic field lines inside the tail, where the IMF is expelled to bend anti-Sunward. For comparison, we also plot the measurement of $B_{\rm{r}}$ from PSP in Figure 8c.  $B_{\rm{r}}$ plotted here has been reduced to its statistical `mode', which is the time sequence with the most probable value in the running intervals of 300 second, which process is implemented to remove most but not all Alfv\'enic turbulence to show the underlying heliospheric magnetic field. We can see that both the simulated and the observed $B_{\rm{r}}$ are mainly negative. On top of the more frequent oscillation of observed $B_{\rm{r}}$, the field variation is close to $B_{\rm{r}}$ (low) as simulated under low cometary activity, while the measured largest spike is coincident with the measured $B_{\rm{r}}$ (high) as derived from simulation under high cometary activity. This comparison, if not coincidence, indicates that PSP crossed the tail between our high-low estimations, maybe caused by 3D motion of the comet tail. The higher frequency variations in the data is due to the broad-band Alfv\'enic fluctuations/spikes/jets that is newly discovered by PSP \citep{Kasper2019, Bale2019}, which is not included in our numerical model. It is unfortunate that, during the third encounter of PSP, the solar wind plasma data is lacked.

In Figure 8d, for the high-level activity case, thermal pressure ($P_{\rm{th}}$) first decreases, then increases and finally recovers to the background level, characterized by a bipolar structure. As the activity weakens, the disturbance of $P_{\rm{th}}$ profile becomes smaller: showing a tiny bump followed by a sharp cliff down to the background level for the middle-level activity, switching to a smooth transition between adjacent thermal pressure levels. Consequently, thermal pressure may not be a good indication of tail crossing for weak comets.

\section{Summary and Discussion}

In this study, we have found the encounter between PSP and a near-Sun comet-like object named 322P. The encounter happened around 21:30 UT on 2019-09-02. During this encounter, PSP was not directly looking at the nucleus of 322P, but might be hit by the dust ejecta. Dynamics of micro-dust particles as released from 322P is simulated and lined up to show the syndynes. The syndynes show remarkable difference before and after the perihelion: they basically follow the trajectory of 322P before the perihelion, and roughly direct away from the Sun for the segment after the perihelion. The trails of dust as viewed from the perspective of PSP are also provided, which illustrates the major part of the trail including 322P itself falls in the mirrored FOV of WISPR while the distant tail may fall in the FOV of WISPR. 

Smaller charged particles (e.g., nanograins, molecular and atomic ions), which are prone to be picked up by the solar wind, are simplified as a single species released from 322P and mass-loaded into the solar wind in our simulation. Due to mass loading, the radial velocity in ambient solar wind declines and westward lateral velocity appears along with the increase of mass density in the plasma. The degree of perturbation is controlled by the activity level. The different spatial gradients between background velocity and disturbed velocity can lead to the formation of shock in front of the mass-loaded plasma structure and rarefaction on the west wing of the structure, if the activity is strong enough. The magnetic field lines as convected by the flow are also deformed by the disturbed velocity: pile up in front of the structure, change the direction and even switch back within the structure for high-level activity. 

We have identified the time of tail crossing for PSP and simulated the time sequences of different variables. The 3D global environment surrounding local disturbances also provides a reasonable context that is helpful to the data interpretation if some signal of mass-loading process can be caught by PSP measurements. By comparing the time series of the simulated and observed magnetic fields, we find that there are some similarities between the simulated and observed $B_{\rm{r}}$ components: (1) the polarity of the $B_{\rm{r}}$ baseline is negative (i.e. towards the sun); (2) $B_{\rm{r}}$ is significantly disturbed or even reversed. However, the observed $B_{\rm{r}}$ disturbance is more frequent, which may be related to the phenomenon of Alfv\'enic spikes. By looking at the electric data from FIELDS/PSP, we find that, on Sep-02 2019, there are about 1353 events with potential difference between antenna tips being larger than 50 mV. Part of these events may be caused by dust impacts, which needs further investigation in future efforts. The small and tiny interplanetary field enhancements (IFEs) will also be another interesting type of structures to be searched carefully, in order to study the interaction between near-Sun solar wind and newly formed dust cloud \citep{Lai2015}. We note that our simple model used here is only able to give a qualitative result to approximate the date of approach for a timely guidance to survey and analyze the PSP data. More accurate models that account for the dynamics of multi-fluid plasmas \citep{Jia2014} shall be applied for a quantitative study once the perturbations are found in the PSP data. 

In summary, our study reports an encounter between a near-Sun small object and PSP, and demonstrate that the available observations are not contradicting our model predictions. We confirm that, according to our modeling results and compared to the in situ measurements, the mass release rate of dusty plasma from 322P comet cannot exceed $2\times10^3$~kg/s or even a smaller upper threshold, indicating that 322P may be on the way to becoming a ``rocky comet''. Without this work, it would be not easy to quantitatively know why PSP misses the main signatures of 322P’s activity. This work also points out the importance of having dual inner-heliosphere imagers looking at both sides of the Sun to catch the rare and valuable opportunity of closeup observation of near-Sun comet’s activity. The same method can be applied to future PSP orbits seeking for more encounters. As the recently launched Solar Orbiter has passed approximately distant downstream of comet Atlas in early June 2020, we expect more cases being observed to put more constraints to our model. Combined with our succeeding quantitative study, will also improve our understanding of the origin, evolution, and fate of the near-Sun comet-like object or active asteroid.

\bigbreak

\noindent Acknowledgements:

The ephemerides of PSP and 322P are used from the JPL/Horizons database. The work at Peking University is supported by NSFC under 41874200, 41421003, 41674171, as well as by CNSA under D020301 and D020302. The work at NSSC, CAS is supported by NSFC under 41774157, 41731067, and 41874202. The MHD simulation part was carried out on TianHe-1(A) at National Supercomputer Center in Tianjin, China. The FIELDS experiment on the Parker Solar Probe spacecraft was designed and developed under NASA contract NNN06AA01C. The FIELDS team acknowledges the contributions of the Parker Solar Probe mission operations and spacecraft engineering teams at the Johns Hopkins University Applied Physics Laboratory. J.-S. He also thank S.-T. Fan, C.-L. Shen, and A. Vourlidas for their kind helps for this work. 

% \begin{figure}[H]
\begin{FPfigure}
%d\begin{figure}
  \centering
    \includegraphics[width=16cm,clip=]{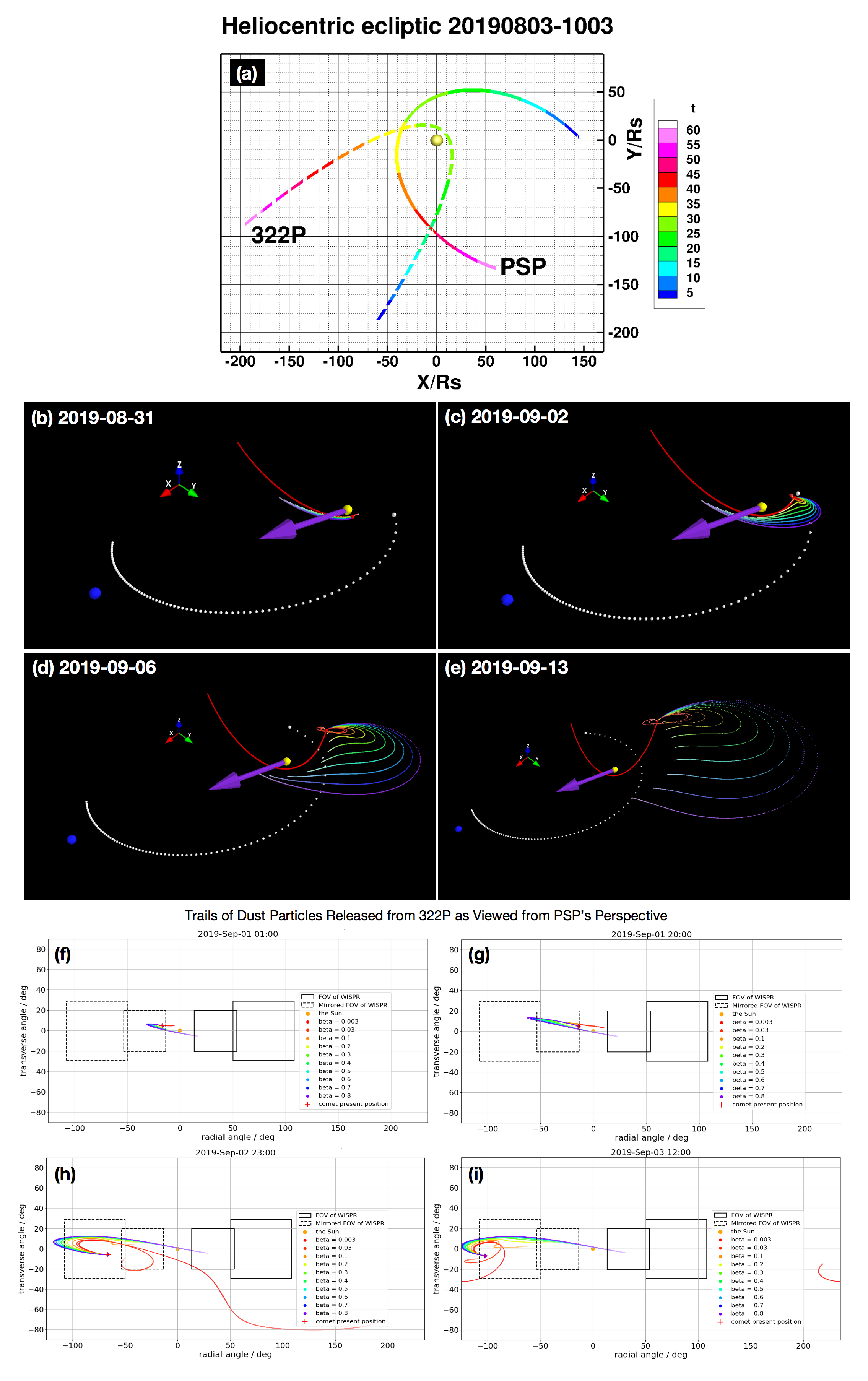}
  \caption{Ephemerides of PSP and 322P as well as the evolution of simulated dust trails. (a) Ephemerides of PSP and 322P in the Heliocentric Ecliptic Coordinate system from 3 August to 3 October. (b-e) Syndynes lines with rainbow colors from red to purple for dust particles of different sizes and $\beta$ values ($\beta$=0.003, 0.03, 0.1, 0.2, 0.4, 0.6, and 0.8). Dust particles on the same syndyne but having smaller saturation values in the HSV color model (close to white color) represent the release from 322P in earlier days (60 days at most in display). The yellow and blue balls are placed at the positions of the Sun and the Earth, respectively. The sequences of small red and white balls represent the trajectories of 322P and PSP. (f-i) Syndynes lines of dust particles with various $\beta$ values as viewed from the perspective of PSP. The FOV of WISPR is plotted with two solid black rectangles, while the mirrored FOV is plotted with two dashed black rectangles. The 322P’s position is marked with a red cross, which mainly moves in the mirrored FOV. }\label{Fig1}
%\end{figure}
\end{FPfigure}

\begin{figure}[htbp]
  \begin{center}
    \includegraphics[width=\textwidth]{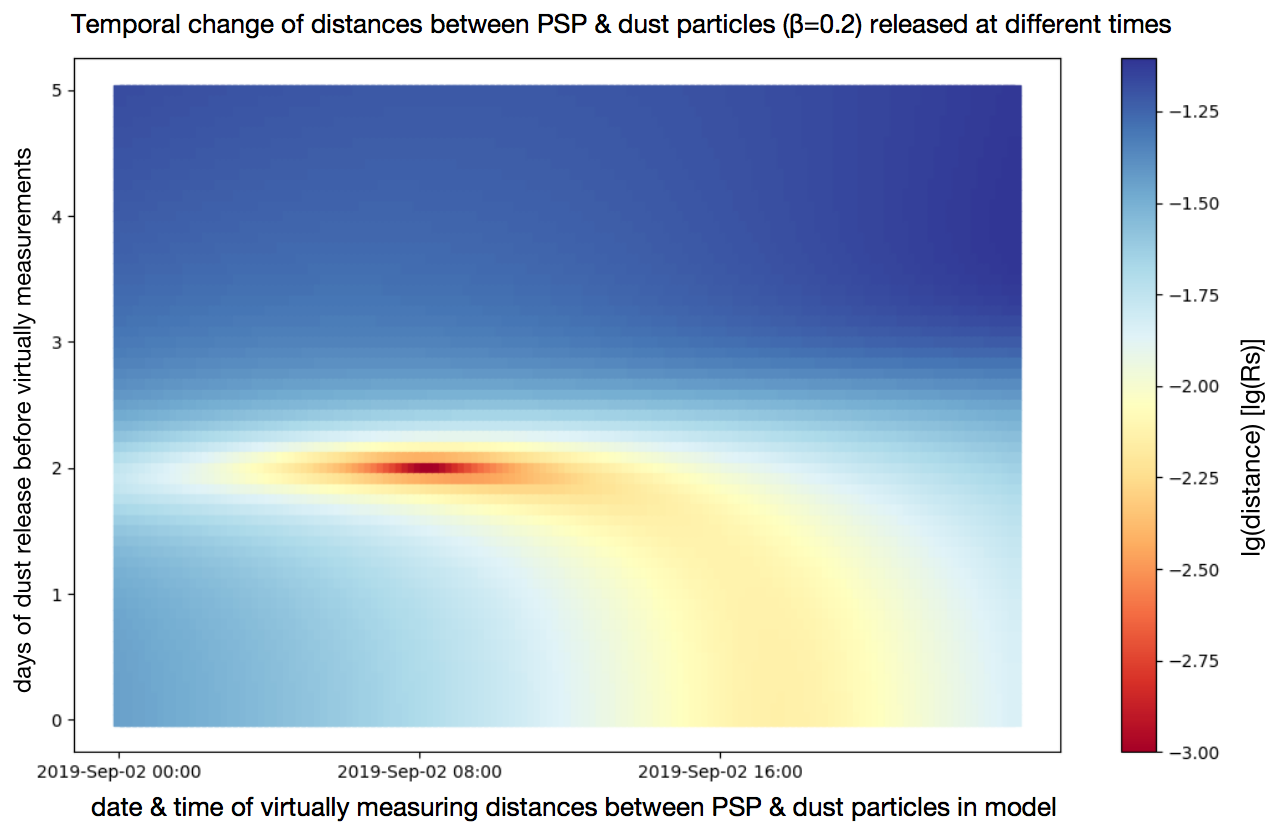}
  \end{center}
  \caption{
  Temporal variation of distances between PSP and simulated dust particles, which are released at different times.}\label{Fig2}
\end{figure}

\begin{figure}[htbp]
  \begin{center}
    \includegraphics[width=\textwidth]{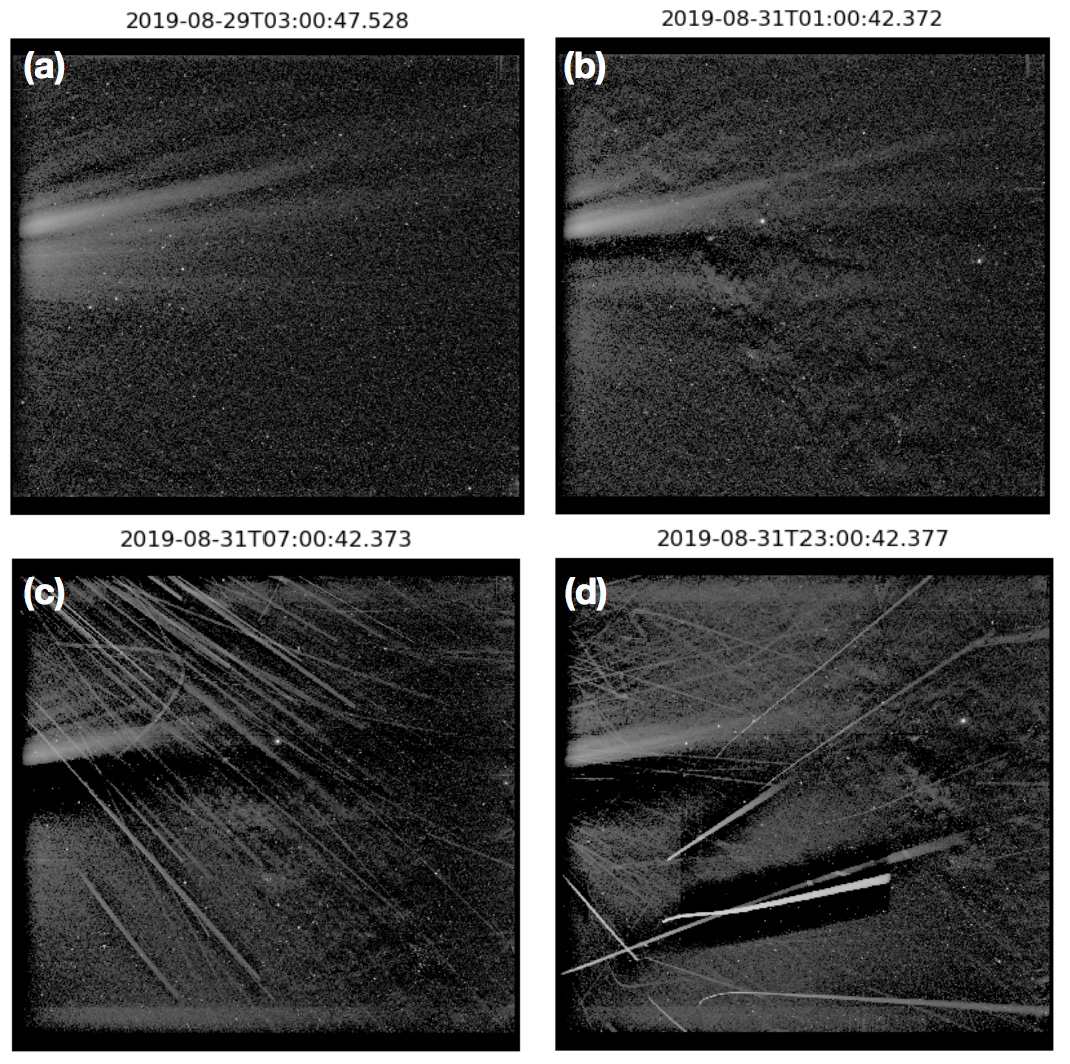}
  \end{center}
  \caption{
    Images photographed by WISPR at 03:00 on August 29 (a), 01:00 on August 31 (b), 07:00 on August 31 (c), and 23:00 on August 31 (d).  
  }\label{Fig3}
\end{figure}

\begin{figure}[htbp]
  \begin{center}
    \includegraphics[width=\textwidth]{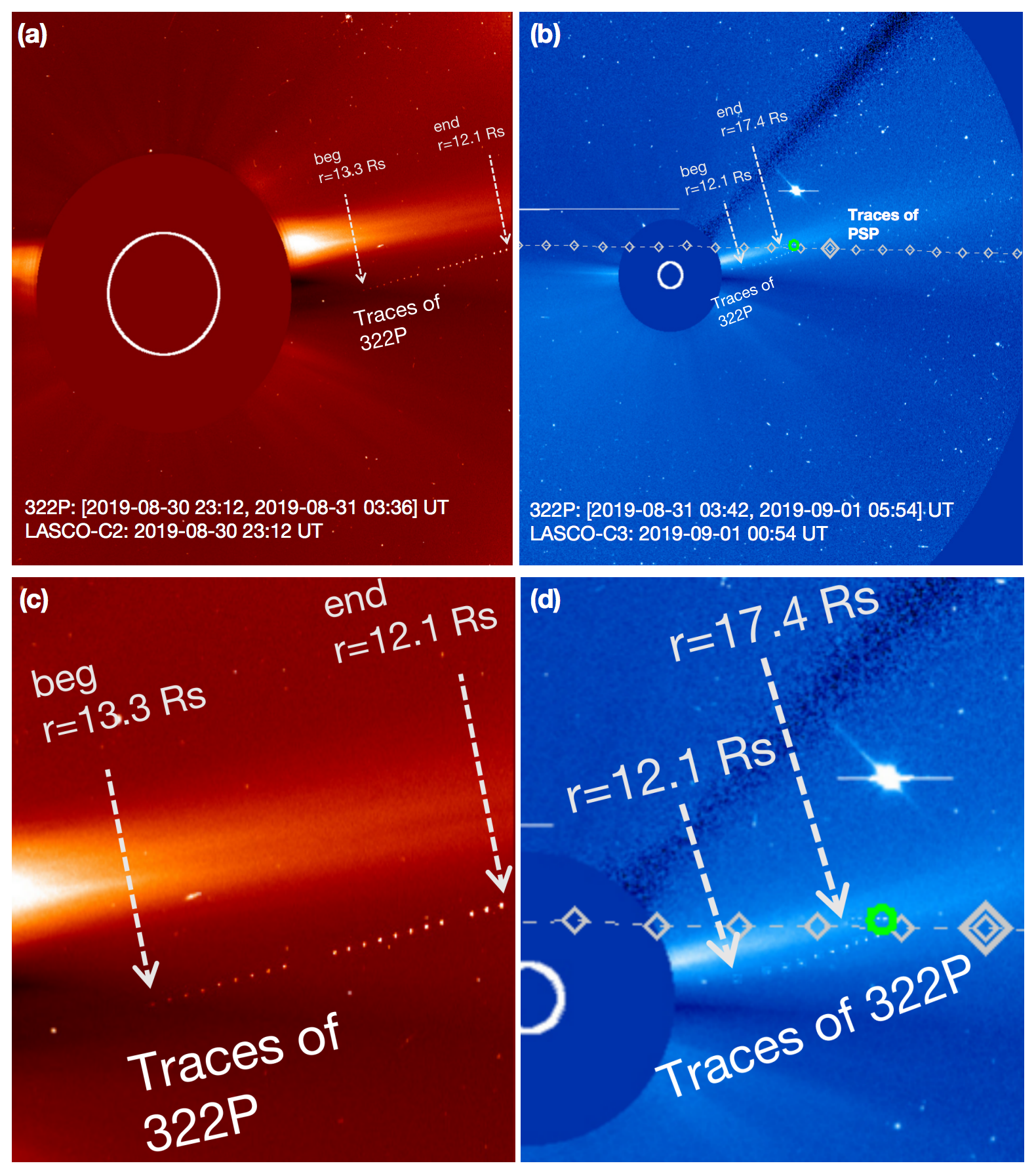}
  \end{center}
  \caption{
    (a) Traces of 322P before its perihelion as observed by LASCO-C2 from 23:12 UT on 2019-08-30 to 03:36 UT on 2019-08-31. As time elapsed, 322P moved rightward in the FOV of C2. (b) Traces of 322P (weak bright dots recorded by LASCO-C3) and PSP (gray diamonds) in the interval of [2019-08-31 03:42, 2019-09-01 05:54] UT as superposed on the background image of LASCO-C3 taken at 00:54 UT on 2019-09-01, when the positions of 322P and PSP are marked with a green circle and a larger gray diamond. (c) Zoom-in of cropped LASCO-C2 image. (d) Zoom-in of cropped LASCO-C3 image.
  }\label{Fig4}
\end{figure}

\begin{figure}[htbp]
  \begin{center}
    \includegraphics[width=\textwidth]{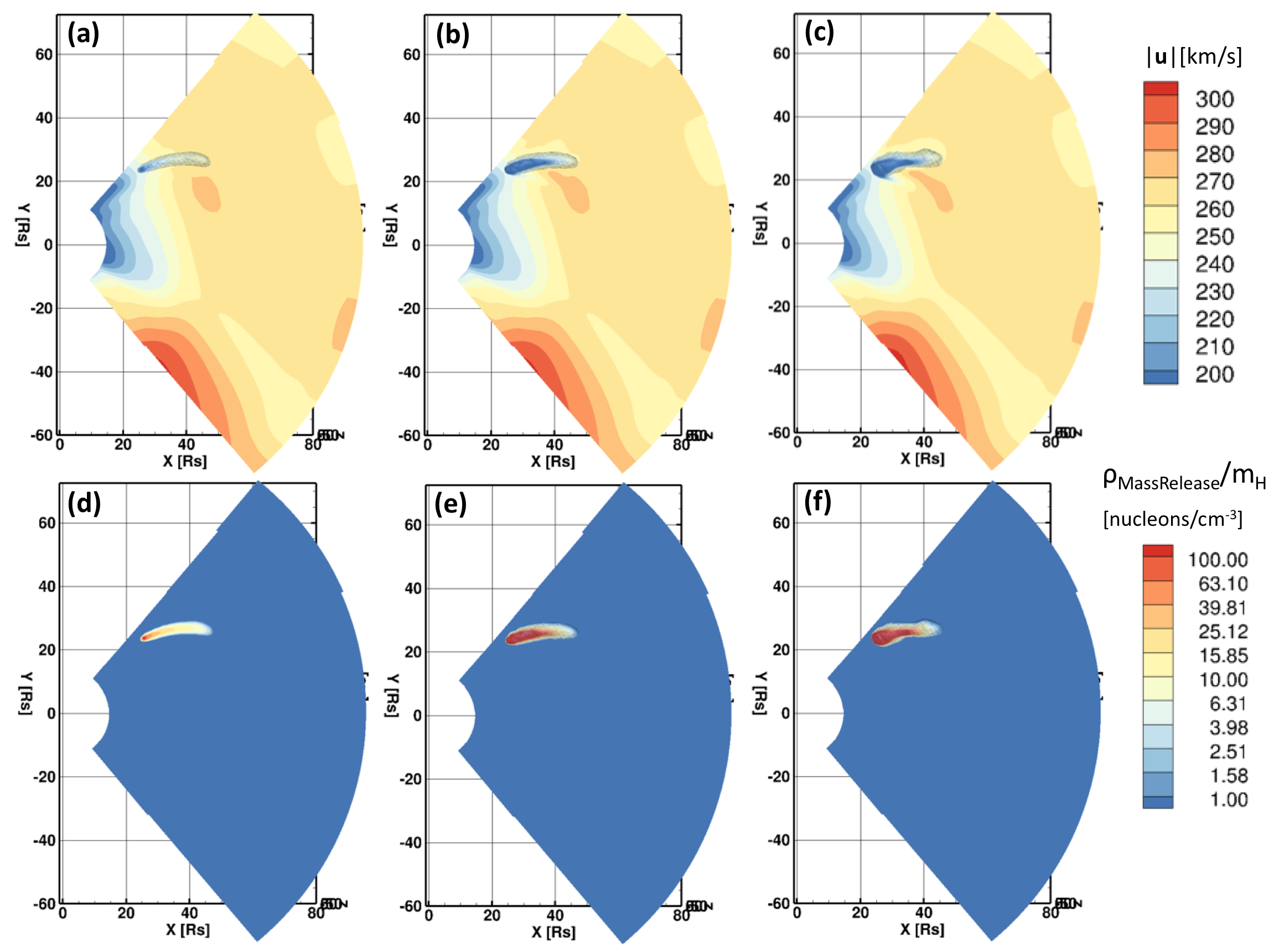}
  \end{center}
  \caption{
Distribution of the disturbed/undisturbed velocity (unit: km/s) (upper panels) and dusty plasma number density in unit of $\rm{amu/cm^3}$ as released from 322P and mass-loaded into the solar wind (lower panels) for the three different levels of activities (left: low, middle: intermediate, right: high) at 09:47 UT on 2019-09-02.
  }\label{Fig5}
\end{figure}

\begin{figure}[htbp]
  \begin{center}
    \includegraphics[width=\textwidth]{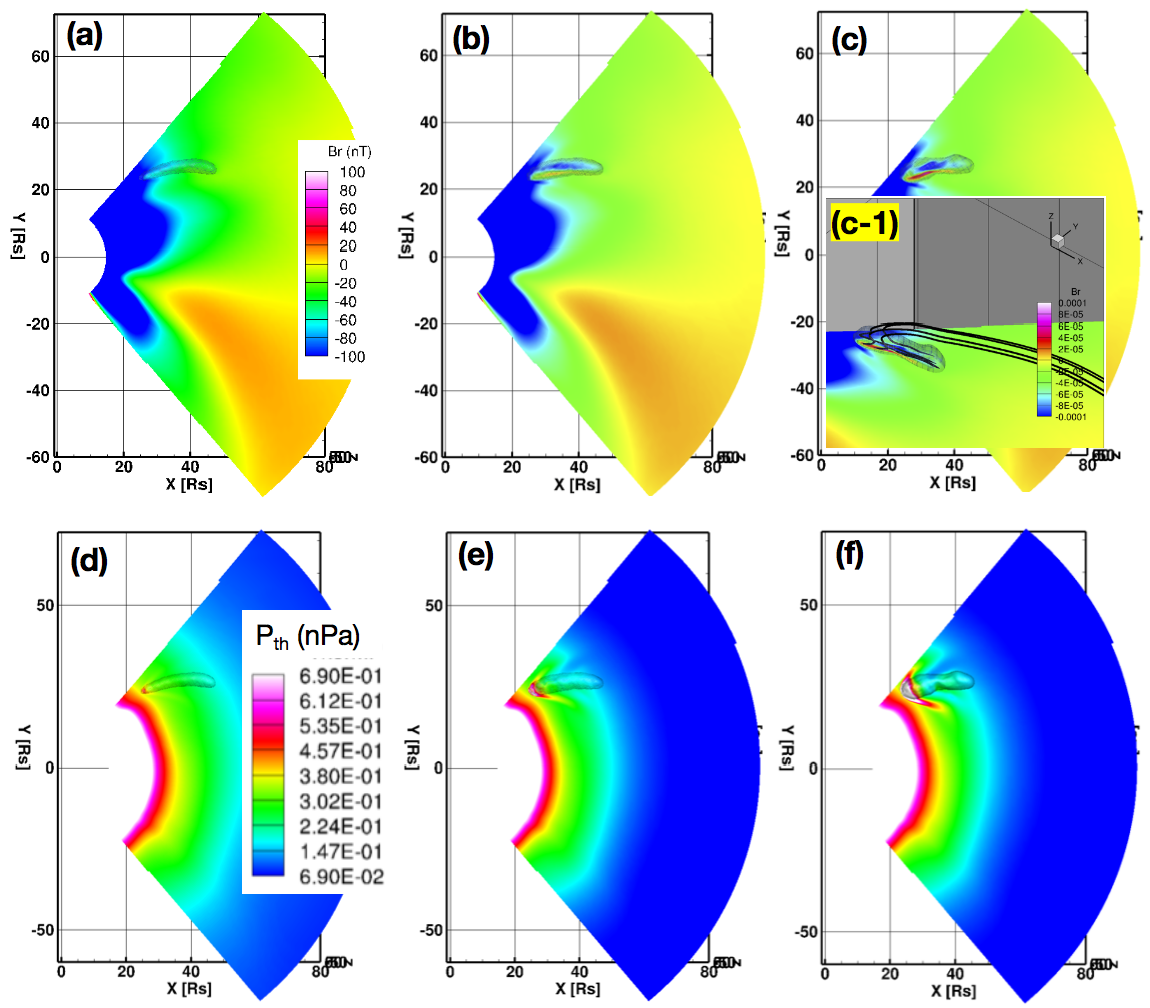}
  \end{center}
  \caption{
Spatial distribution of $B_{\rm{r}}$ (upper panels: a-c) and $P_{\rm{th}}$ (lower panels: d-f) near and within the mass-loaded (dusty plasma) structure. Enhanced $B_{\rm{r}}$ (negative) and $P_{\rm{th}}$ can be found downstream of the shock region, the formation of which is obvious in Panels (b) \& (c) \& (e) \& (f) for middle and high levels of activities. Disturbed magnetic field lines showing switch-back / kinked geometry are displayed in the inset panel (c-1). Weakened $P_{\rm{th}}$, a signature of rarefaction region, appears on the west side of the mass-loaded structure in Panels (e) \& (f). 
  }\label{Fig6}
\end{figure}

\begin{figure}[htbp]
  \begin{center}
    \includegraphics[width=\textwidth]{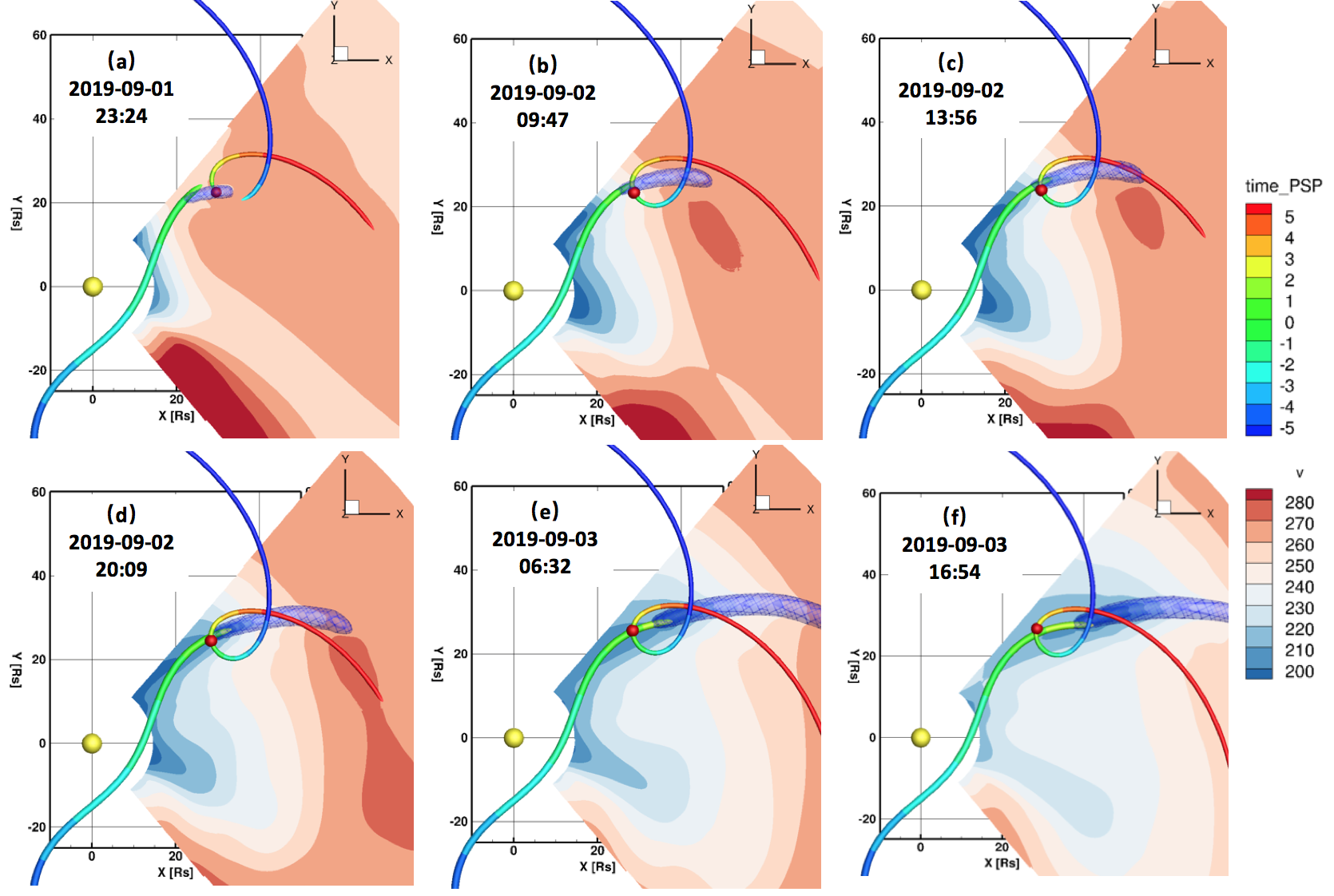}
  \end{center}
  \caption{
Movements of PSP (red solid ball) and 322P (head of the blue isosurface) along their trajectories (solid threads color coded with the days (unit: day) centered around 00:00 UT on 2019-09-02 using the rainbow table) during their encounter in the Carrington coordinate system. The blue isosurface ($\rho_{\rm{MassRelease}}$=1~$\rm{amu/cm^3}$ ) shows the mass-loaded (dusty plasma) structure. The background shows the plasma speed contour sliced in the plane defined by three points: the Sun, the object nucleus, and the end of dusty plasma tail contour.
  }\label{Fig7}
\end{figure}

\begin{figure}[htbp]
  %\begin{center}
    %\includegraphics[width=\textwidth]{Figure 6 (2020-01-24).png}
    \centerline{\includegraphics[width=\textwidth,clip=]{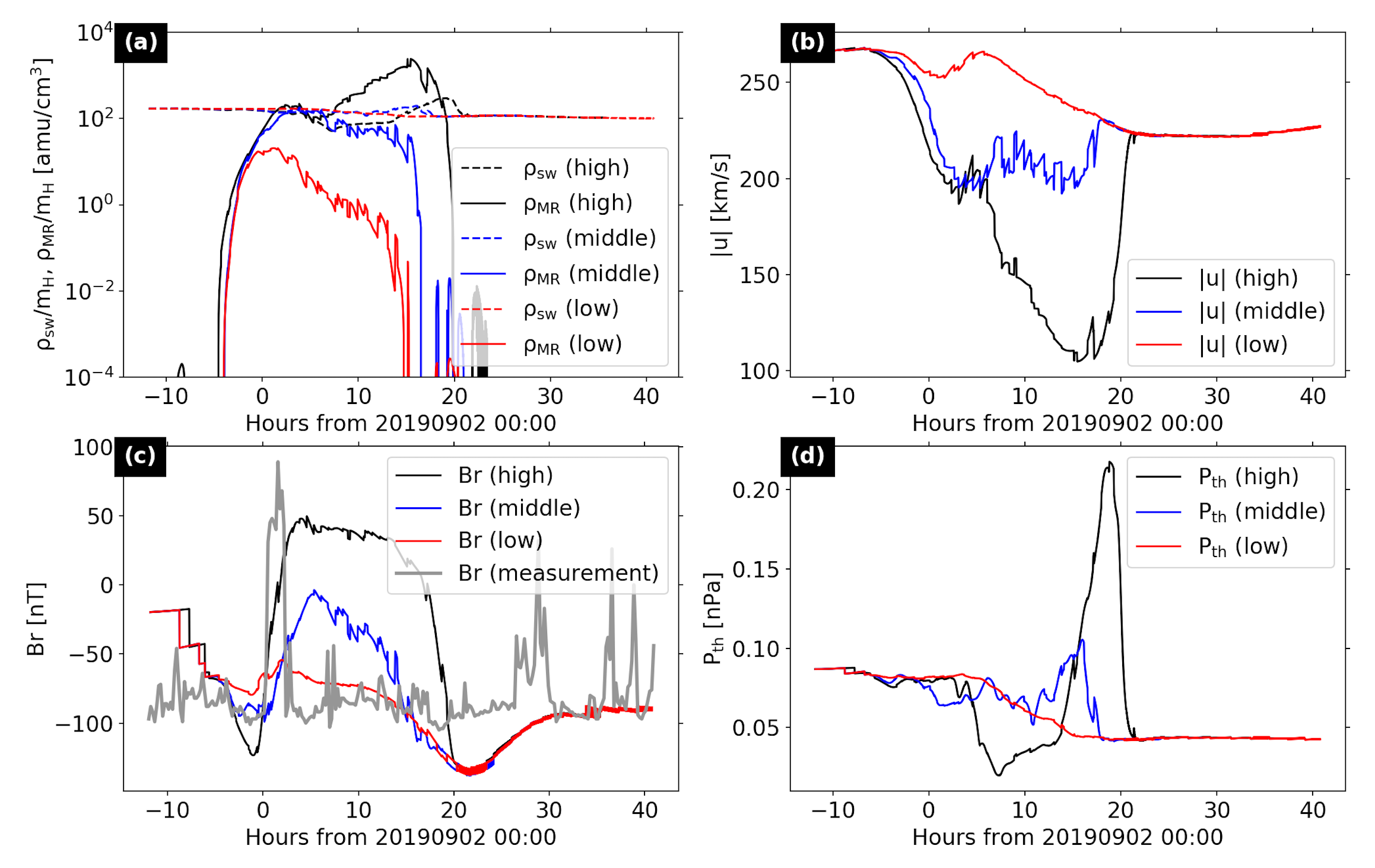}}
  %\end{center}
  \caption{
Variables sampled along the path of PSP during its encounter with 322P and cross of the mass-loaded plasma structure. The variables for the low, middle, and high levels of activities are plotted in red, blue, and black, respectively. (a) Densities of released cometary plasma (solid lines) and mass-loaded solar wind plasma (dashed lines). (b) Velocity of the mass-loaded plasma. (c) Radial component of disturbed interplanetary magnetic field vector. Statistical `mode' of measurement of $B_{\rm{r}}$ from PSP/FIELDS is plotted in light gray color for comparison.  (d) Thermal pressure of mass-loaded plasma. }\label{Fig8}
\end{figure}

%\begin{figure}[htb!]
%	\centerline{\includegraphics[width=18cm,clip=]{Figure6new.eps}}
%	\caption{Construction results based on fitting parameters $\mu(\tau)$ and $\lambda^2(\tau)$ of Castaing theoretical expressions (red lines) and observational results (black lines) for structure functions from order 0.5 to 6 with an increment of 0.5 (\textit{Left Column}), scaling exponents (\textit{Middle Column}) and flatness (\textit{Right Column}). Datasets are from FGM (upper panels), SCM (middle panels) and EDP (bottom panels), respectively. (All the results are for Event 7)}
%	\label{fig6}
%\end{figure}

\bibliographystyle{aasjournal}
\bibliography{References_for_PSP_322P}

\end{document}